\documentclass[sigconf]{acmart}

\usepackage{booktabs} 

\usepackage[T1]{fontenc}

\usepackage{amsfonts}
\usepackage{dsfont}

\makeatletter
\def\set@curr@file#1{%
  \begingroup
    \escapechar\m@ne
    \xdef\@curr@file{\expandafter\string\csname #1\endcsname}%
  \endgroup
}
\def\quote@name#1{"\quote@@name#1\@gobble""}
\def\quote@@name#1"{#1\quote@@name}
\def\unquote@name#1{\quote@@name#1\@gobble"}
\makeatother

\usepackage{graphicx}

\newcommand{\kww}[1]{{\color{blue}\texttt#1}}

\newcommand{\holl}{\textsf{HOL Light}}

\newtheorem{definition}{Definition}[section]

\newtheorem{theorem}{Theorem}[section]

\copyrightyear{2020}
\acmYear{2020}
\setcopyright{acmcopyright}
\acmConference[SAC '20]{The 35th ACM/SIGAPP Symposium on Applied Computing}{March 30-April 3, 2020}{Brno, Czech Republic}
\acmBooktitle{The 35th ACM/SIGAPP Symposium on Applied Computing (SAC '20), March 30-April 3, 2020, Brno, Czech Republic}
\acmPrice{15.00}
\acmDOI{10.1145/3341105.3373993}
\acmISBN{978-1-4503-6866-7/20/03}

\begin{document}
\title{Formal Analysis of the Biological Circuits using Higher-order-logic Theorem Proving}

\author{Sa'ed Abed}
\orcid{1234-5678-9012}
\affiliation{%
  \institution{Computer Engineering Department \\ College of Engineering and Petroleum, Kuwait University}
  \city{Kuwait}
}
\email{s.abed@ku.edu.kw}
\author{Adnan Rashid}
\affiliation{%
  \institution{School of EE and CS (SEECS) \\ National University of Sciences and Technology (NUST)}
  \city{Islamabad}
  \country{Pakistan}}
\email{adnan.rashid@seecs.nust.edu.pk}
\author{Osman Hasan}
\affiliation{%
  \institution{School of EE and CS (SEECS) \\ National University of Sciences and Technology (NUST)}
  \city{Islamabad}
  \country{Pakistan}}
\email{osman.hasan@seecs.nust.edu.pk}

\renewcommand{\shortauthors}{S. Abed et al.}


\begin{abstract}

Synthetic Biology is an interdisciplinary field that utilizes well-established engineering principles, ranging from electrical, control and computer systems, for analyzing the biological systems, such as biological circuits, enzymes, pathways and controllers.
Traditionally, these biological systems, i.e., the genetic circuits are analyzed using paper-and-pencil proofs and computer-based simulations techniques. However, these methods cannot provide accurate results due to their inherent limitations such as human error-proneness, round-off errors and the unverified algorithms present in the core of the tools, providing such analyses. In this paper, we propose to use higher-order-logic theorem proving as a complementary technique for analyzing these systems and thus overcome the above-mentioned issues.
In particular, we propose a higher-order-logic theorem proving based framework to formally reason about the genetic circuits used in synthetic biology. The main idea is to, first, model the continuous dynamics of the genetic circuits using differential equations. The next step is to obtain the systems' transfer function from their corresponding block diagram representations. Finally, the transfer function based analysis of these differential equation based models is performed using the Laplace transform. To illustrate the practical utilization of our proposed framework, we formally analyze the genetic circuits of activated and repressed expressions of protein.
\end{abstract}

%
%



\maketitle



\section{Introduction}\label{SEC:intro}

Nowadays, engineering principles~\cite{muschler2004engineering} are being widely adapted in analyzing biological systems~\cite{haefner2005modeling}, signaling pathways~\cite{alberts2017molebiocell} (a group of molecules working together to control various functions of a cell) and biological circuits~\cite{baldwin2012synthbio} (the biological parts of a cell mimicking the logical functionality that is performed in electrical circuits) etc. For example, control systems laws~\cite{ogata2002modern} are used for analyzing biological systems, i.e., genetic circuits and bio-controllers. The amalgamation of such interdisciplinary fields into synthetic biology~\cite{baldwin2012synthbio} allows designing and analyzing these systems in an efficient manner.

The analysis of these systems, i.e., biological circuits and bio-controllers, requires modeling their dynamics, representing the interaction of their different components, using differential equations. Next, the transfer functions of these systems, providing their dynamics in the frequency domain are extracted from their block diagram representations~\cite{nise2007control}, which are commonly used in control systems. Finally, the Laplace transform is used to perform the frequency domain (transfer function based) analysis of these systems based on their differential equation models.

Traditionally, these biological systems, i.e., biological circuits, networks and pathways, are analyzed using the paper-and-pencil proof~\cite{de1982evaluation} and computer based numerical~\cite{gaylord1996computer} and simulation methods~\cite{haefner2005modeling}. However, the paper-and-pencil proof methods are prone to human error and the chances of errors increase while analyzing the larger systems. Similarly, the computer-based numerical techniques are based on the unverified numerical algorithms that are present in the core of the associated tools. Also, the simulation-based methods suffer from the limited computational resources and computer memory issues. Thus, considering the safety-critical nature of biological systems, these conventional methods cannot be relied upon for their accurate analysis.

\begin{figure*}[!ht]
\centering
\scalebox{0.270}
{\hspace*{-3.0cm} \includegraphics[trim={5.0 0.4cm 5.0 0.4cm},clip]{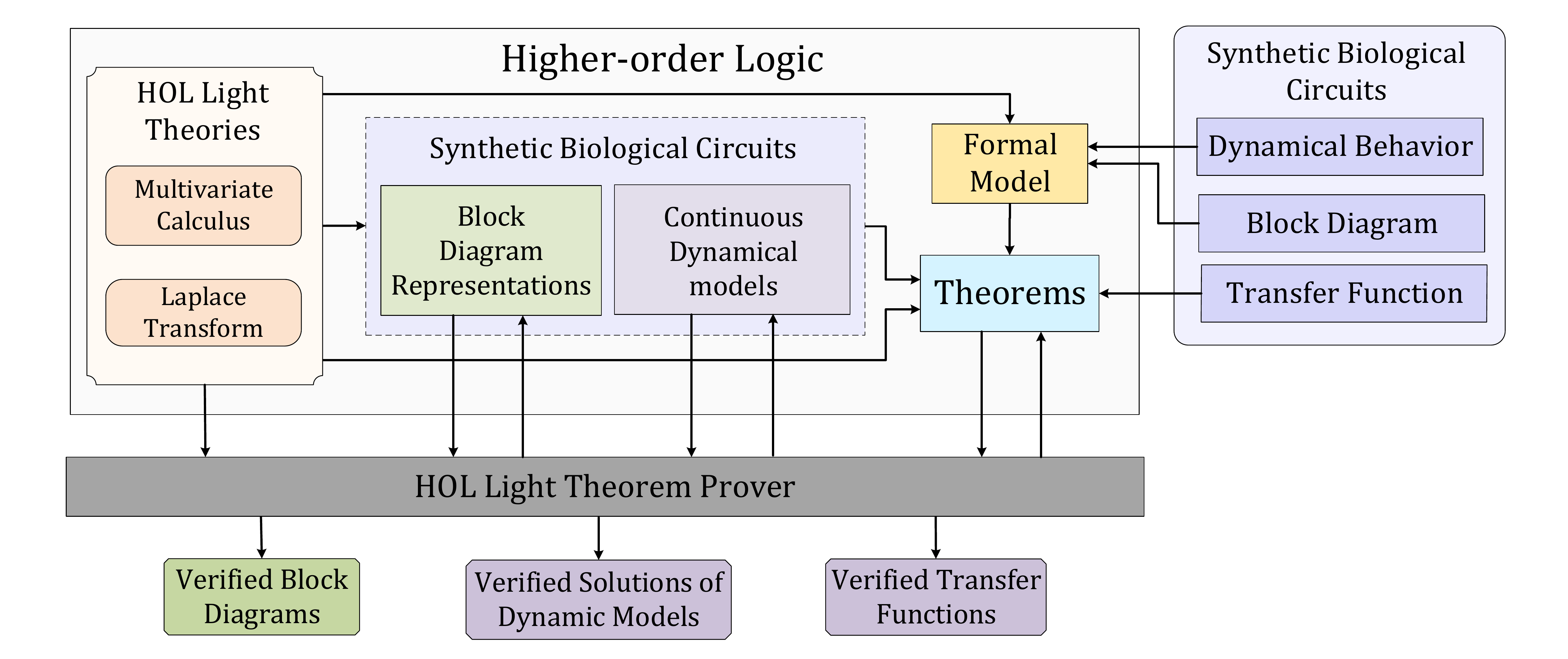}}
\caption{Proposed Framework}
\label{FIG:proposed_framework}
\end{figure*}

Formal methods~\cite{hasan2015formal} are computer-based mathematical techniques used for the modeling, specification and verification of the systems. They have been widely adopted for the rigorous analysis of the complex real-world systems~\cite{rashid2017formal,camilleri1986hardware}. They are mainly of two types, i.e., model checking~\cite{hasan2015formal} and theorem proving~\cite{hasan2015formal}. Model checking involves the development of a state-space model of the underlying system and the verification of its intended behavior by the properties specification in an appropriate logic.
It has been extensively used in the area of synthetic biology for formally analyzing the biological circuits and their associated feedback controllers~\cite{yordanov2012experimentally,bartocci2013temporal,madsen2012utilizing}.
However, it suffers from the inherent state-space explosion problem~\cite{clarke2001progress} and thus is not suitable for analyzing larger systems.
On the other hand, higher-order-logic theorem proving involves constructing a mathematical model of the system based on higher-order logic and verification of its intended properties using deductive reasoning.
It has been used to formal reason about the biological systems and molecular pathways. Ahmad el at.~\cite{ahmad2014formalization} proposed a formalization of Zsyntax using HOL theorem prover. Moreover, they utilized their proposed framework for formally analyzing the TP53 degradation pathway and Glycolytic leading from D-Glucose to Fructose-1,6-bisphosphate. Recently, Rashid et al.~\cite{rashid2017formal} developed a framework, using the \holl~theorem prover, which provides the formal support for the reaction kinetic based dynamical analysis of the biological systems. However, their proposed framework cannot model and analyze the genetic circuits.

In this paper, we propose to use higher-order-logic theorem proving~\cite{hasan2015formal} for formally analyzing the genetic circuits used in synthetic biology as shown in Figure~\ref{FIG:proposed_framework}. The first step is to model the continuous dynamics of the genetic circuits using differential equations. The next step requires the transfer function of these systems that can be obtained from their corresponding block diagram representations. Finally, the Laplace transform is used to perform the transfer function based analysis of the differential equation based models of these circuits. To illustrate the practical utilization of our proposed framework, we formally analyze the genetic circuits of activated and repressed expressions of protein using \holl.

The rest of the paper is organized as follows: We provide an introduction about the \holl~theorem prover, multivariate calculus and the Laplace transform theories of \holl~in Section~\ref{SEC:prelim}. Section~\ref{SEC:form_block_diagram_rep} presents the formalization of block diagram representations of the biological circuits. We provide our formal analysis of the genetic circuits of activated and repressed expressions of protein using \holl~in Section~\ref{SEC:formal_analysis_genetic_circuits}. Finally, Section~\ref{SEC:summary} concludes the paper.


\section{Preliminaries} \label{SEC:prelim}
This section presents a brief introduction to the \holl~proof assistant and its multivariate calculus and the Laplace transform theories, which are extensively used in the rest of the paper.

\subsection{Theorem Proving and \holl~Theorem Prover}\label{SUBSEC:tp_hol_light_tp}

Theorem proving is a widely used formal verification method that involves developing the proofs of the mathematical theorems using a computer program (called \emph{theorem prover})~\cite{harrison_book}.
Theorem proving systems have been commonly utilized for formally verifying the properties of both hardware and software systems~\cite{camilleri1986hardware,schumann2001automated}. Based on the expressiveness requirement, these properties are modeled as theorems using propositional, first-order or higher-order logic. For example, the higher-order logic provides more expressiveness by allowing additional quantifiers. Moreover, it is best suited for conducting the mathematical analysis based on theories of multivariate calculus and the Laplace transform.

\holl~\cite{harrison1996hol} is an interactive theorem prover for developing the formal proofs of the mathematical concepts expressed in the form of theorems. It is implemented in Objective CAML (OCaml), which is a functional programming language, with an aim of automating the mathematical proofs~\cite{Ocaml_ref}. It has a very small logical kernel of approximately $400$ lines of OCaml code and has types, terms, axioms, inference rules and theorems as its main components, which are a part of its theory. Every new theorem in a theory is verified using the basic axioms and the primitive inference rules or already verified theorems, providing the soundness of this method.


\begin{table*}[!ht]
    \caption{Laplace Transform}
    \label{TAB:lap_trans_for}
    \begin{tabular}{ |p{4.0cm}|p{9.0cm}| }
        \hline 
        Mathematical Form & Formalized Form   \\  \hline \hline
        \multicolumn{2}{|c|}{\textbf{Laplace Transform}} \\ \hline
        { {$\begin{array} {lcl} \mathcal{L} [f(t)] = F(s) =  \\
            \hspace*{0.2cm}  \int_{0}^{\infty} {f(t)e^{-s t}} dt, \ s \ \epsilon \  \mathds{C}
                \end{array}$}  }
        &
        {$\begin{array} {lcl} \textup{\texttt{  \hspace*{-0.4cm} $\vdash_{\mathit{def}}$ $\forall$s f. \kww{laplace\_transform} f s =    }} \\
            \textup{\texttt{   \hspace*{0.4cm}  integral \{t $|$ \&0 $<=$ drop t\}  }}  \\
            \textup{\texttt{   \hspace*{0.9cm}  ($\lambda$t. cexp ($--$(s $\ast$ Cx (drop t))) $\ast$ f t) \hspace*{0.1cm}  }}
            \end{array}$}  \\ \hline
        \multicolumn{2}{|c|}{\textbf{Laplace Existence}} \\ \hline
        \vspace*{-0.5cm}
        $f$ is piecewise smooth and is of exponential order on the positive real line
        &
        {$\begin{array} {lcl} \textup{\texttt{  \hspace*{-0.4cm} $\vdash_{\mathit{def}}$ $\forall$s f. \kww{laplace\_exists} f s $\Leftrightarrow$  }} \\
            \textup{\texttt{  \hspace*{0.2cm} ($\forall$b. f piecewise\_differentiable\_on  \hspace*{0.1cm} }}  \\
            \textup{\texttt{ \hspace*{2.0cm}  interval [lift (\&0),lift b]) $\wedge$   }}  \\
            \textup{\texttt{  \hspace*{0.2cm}  ($\exists$M a. Re s $>$ drop a $\wedge$ exp\_order\_cond f M a) \hspace*{0.1cm} }}
            \end{array}$}  \\ \hline
        \multicolumn{2}{|c|}{\textbf{Exponential-order Condition}} \\ \hline
        \vspace*{-0.55cm}
        There exist a constant $a$ and a positive constant $M$ such that $|f (t)| \leq Me^{at}$
        &
        {$\begin{array} {lcl} \textup{\texttt{ \hspace*{-0.4cm} $\vdash_{\mathit{def}}$ $\forall$f M a. \kww{exp\_order\_cond} f M a $\Leftrightarrow$ \&0 $<$ M $\wedge$ \hspace*{0.1cm}  }} \\
            \textup{\texttt{ \hspace*{0.0cm}  ($\forall$t. \&0 $<=$ t $\Rightarrow$    }}  \\
            \textup{\texttt{ \hspace*{0.7cm}  ||f (lift t)|| $<=$ M $\ast$ exp (drop a $\ast$ t)) \hspace*{0.1cm} }}
            \end{array}$}  \\ \hline
    \end{tabular}
\end{table*}


\subsection{Multivariable Calculus and Laplace Transform Theories}

\holl~presents an extensive support for analyzing the systems using theories of multivariable calculus and the Laplace transform. Table~\ref{TAB:lap_trans_for} provides some definitions from ~\holl's theory of the Laplace transform, which includes the Laplace transform, Laplace existence and the exponential-order conditions. Interested readers can refer to~\cite{taqdees2013formalization,rashid2017tmformalization,adnan2018JAL,rashid2017formal} for more details about this theory. It is extensively utilized in our formal analysis of the biological circuits.


\section{Formalization of Block Diagram Representations of Biological Circuits}\label{SEC:form_block_diagram_rep}

In this section, we present our formalization of block diagram representations of the biological circuits. These definitions enable us to formally model the block diagrams of a generic biological circuit in the s-domain and thus to find out the transfer function of any biological circuit. The presented formalization is basically inspired from the block diagrams of the control systems~\cite{houpis2013linear,ahmad2014formal}.

Configuration $1$: The transfer function of a sum of two components (subsystems) of a biological circuit, which can be any proteins or genes, in the case of a genetic circuit, is equal to the product of the transfer function of the individual components.
We formalize this configuration for an arbitrary (\textit{N}) number of components of a circuit as follows:

\begin{definition}
\label{DEF:series_rep_comp}
\emph{Series Components} \\{
\textup{\texttt{
\hspace*{0.0cm} $\vdash_{\mathit{def}}$ $\forall$C$\mathtt{_i}$. \kww{series\_comp} [C$\mathtt{_1}$; C$\mathtt{_2}$; ...; C$\mathtt{_N}$] = $\mathtt{\prod\limits_{i = 1}^{N}}$ C$\mathtt{_i}$
}}}
\end{definition}

The function \texttt{series\_comp} accepts the transfer functions of the individual components of the circuit as a list of complex numbers and returns the transfer function of the overall circuit as a product of all individual transfer functions.

Configuration $2$: The summation junction for various components of a biological circuit is an addition module that provides the summation of the transfer functions of the individual components.
We formalize this configuration for an arbitrary number (\textit{N}) of components of a circuit, having transfer functions represented by a list of complex numbers as follows:

\begin{definition}
\label{DEF:summation_jun}
\emph{Components as Summation Junction} \\{
\textup{\texttt{
\hspace*{0.0cm} $\vdash_{\mathit{def}}$ $\forall$C$\mathtt{_i}$. \kww{summ\_jun} [C$\mathtt{_1}$; C$\mathtt{_2}$; ...; C$\mathtt{_N}$] = $\mathtt{\sum\limits_{i = 1}^{N}}$ C$\mathtt{_i}$
}}}
\end{definition}

Configuration $3$: The pickoff point configuration is the representation of a component of a biological circuit to a parallel branch of components.
We model this configuration in \holl~as follows:

\begin{definition}
\label{DEF:pickoff_point}
\emph{Components as Pickoff Point} \\{
\textup{\texttt{
$\vdash_{\mathit{def}}$ $\forall \mathtt{\alpha}$ C$\mathtt{_i}$. \kww{pick\_point} [C$\mathtt{_1}$; C$\mathtt{_2}$; ...; C$\mathtt{_N}$] =    \\
}}
\textup{\texttt{
\hspace*{3.42cm} [$\mathtt{\alpha}\ \ast$ C$\mathtt{_1}$; $\mathtt{\alpha}\ \ast$ C$\mathtt{_2}$; ...; $\mathtt{\alpha}\ \ast$ C$\mathtt{_N}$]
}}}
\end{definition}

The function \texttt{pick\_point} accepts the transfer function of the first component as a complex number and the transfer functions of the components in parallel as a list of complex numbers, and returns the corresponding transfer functions corresponding to the equivalent block diagram representation as a list of complex
numbers.

Configuration $4$: The feedback block configuration is the fundamental representation for modeling the closed loop controllers for the biological circuits.
Due to the presence of the feedback signal, it is primarily represented by an infinite summation of branches that consist of serially connected components.
We formalize the transfer function of each branch as the following \holl~function:

\begin{definition}
\label{DEF:tf_branch}
\emph{Transfer Function of a Branch} \\{
\textup{\texttt{
$\vdash_{\mathit{def}}$ $\forall \mathtt{\alpha}\ \mathtt{\beta}$ N. \kww{branch\_tf} $\mathtt{\alpha}\ \mathtt{\beta}$ N =  $\mathtt{\prod\limits_{i = 0}^{N}}$ series\_comp [$\mathtt{\alpha}; \mathtt{\beta}$]
}}}
\end{definition}

The function \texttt{branch\_tf} takes the forward path transfer function $\mathtt{\alpha}$ (a protein or gene), the feedback path (feedback signal) transfer function $\mathtt{\beta}$ and the number of the branch (\textit{N}), and returns a complex number representing the transfer function of the $N^{th}$ branch.

Next, we formalize the feedback block representation using our function \texttt{branch\_tf} as follows:

\begin{definition}
\label{DEF:feedback_block_rep_comp}
\emph{Feedback Block Representation of Components} \\{
\textup{\texttt{
$\vdash_{\mathit{def}}$ $\forall$$\mathtt{\alpha}$ $\mathtt{\beta}$. \kww{feedback\_block} $\mathtt{\alpha}$ $\mathtt{\beta}$ =    \\
}}
\textup{\texttt{
\hspace*{3.42cm} series [$\mathtt{\alpha}$; $\sum\limits_{k = 0}^{\infty}$ branch $\mathtt{\alpha}$ $\mathtt{\beta}$ k]
}}}
\end{definition}

The function \texttt{feedback\_block} accepts the forward path transfer function $\mathtt{\alpha}$ and the feedback path transfer function $\mathtt{\beta}$ and returns the transfer function by forming the series network
of the final forward path transfer function and the summation of all the possible infinite branches.

Our formalization of the foundational configurations, presented above~\cite{adnan19fabcholtp}, enables us to formally model the block diagram representations of the generic biological circuits as will be illustrated in the next section.


\section{Formal Analysis of the Genetic Circuits}\label{SEC:formal_analysis_genetic_circuits}

Gene expression is a technique for transmitting information from the passive deoxyribonucleic acid (DNA) to the active proteins, which are widely used for performing majority of the tasks required for a human life at cellular level. The process of this transmission is performed in two steps. In the first step, a section of DNA is read out into ribonucleic acid (RNA) and is known as transcription. The second step, namely translation, involves conversions of a short strand of transcribed RNA into protein. This process is usually regulated in synthetic systems by transcription factors (TFs), which control the initiation rate of the transcription of a gene and its corresponding expression. TFs are of two types, namely activators and repressors. Activators increase the transcription rate, whereas the repressors inhibit transcription.
We use our proposed formalization for formally analyzing the genetic circuits of the activated and repressed expressions of protein.

\subsection{Activated Expression of Protein}\label{SUBSEC:activ_express_protein}

\begin{figure}[!ht]
\centering
\scalebox{0.420}
{\hspace*{-0.4cm} \includegraphics[trim={5.0 0.4cm 5.0 0.4cm},clip]{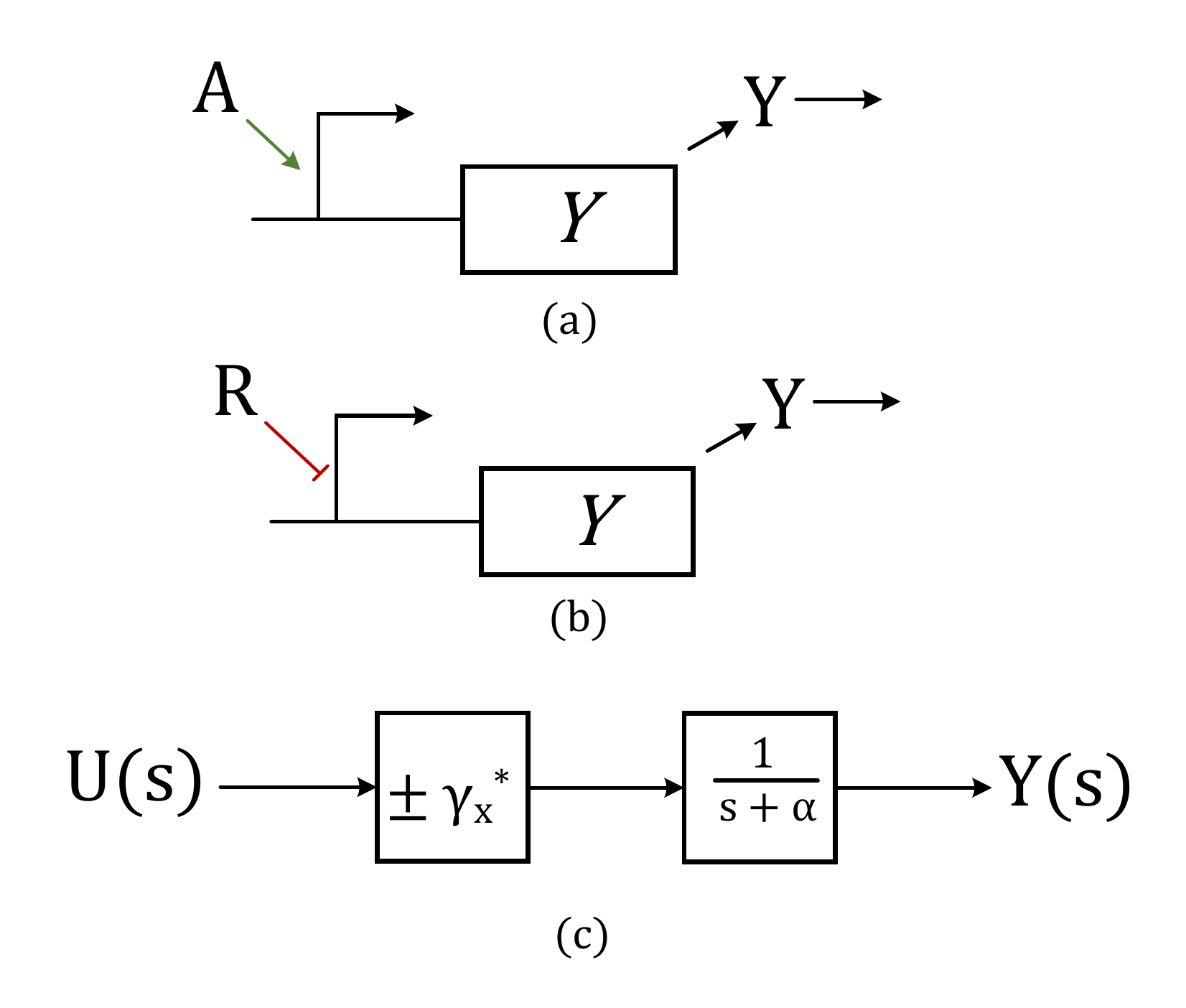}}
\caption{(a) Genetic Circuit Diagram of Activated Expression (b) Genetic Circuit Diagram of Repressed Expression (c) Block Diagram Representation for a Single Gene with Activated and Repressed Expressions}
\label{FIG:genetic_circuit_bd}
\end{figure}

The genetic circuit of the activated expression of protein is depicted in Figure~\ref{FIG:genetic_circuit_bd}(a).  It involves the interaction of the incoming activating TF A with its promoter and the
regulation of the expression of gene $Y$ producing the protein Y. The block diagram representation of the activated expression is shown in Figure~\ref{FIG:genetic_circuit_bd}(c), by a gain block of $+\gamma_x^{\ast}$ with $x = A$.

The dynamical model of the activated expression of protein is mathematically expressed as the following linear differential equation:

\begin{equation}\label{EQ:dyn_mod_activ_expr}
\frac{dy}{dt} + \alpha y = \gamma_A^{\ast} u
\end{equation}

The corresponding transfer function is represented as:

\begin{equation}\label{EQ:trans_fun_activ_expr}
\frac{Y (s)}{U (s)} = \frac{\gamma_A^{\ast}}{s + \alpha}
\end{equation}

In order to model the dynamical behaviour, we first model the linear differential equation of order $n$ as:

\begin{definition}
\label{DEF:diff_eq_order_n}
\emph{Differential Equation of Order $n$} \\{
\textup{\texttt{
$\vdash_{\mathit{def}}$ $\forall$n f t. \kww{differ\_equat\_order\_n} lst f t =    \\
}}
\textup{\texttt{
\hspace*{2.00cm} vsum (0..n) ($\lambda$k. EL k [$\mathtt{\alpha_1}$;$\mathtt{\alpha_2}$;...;$\mathtt{\alpha_k}$] $\ast$
}} \\
\textup{\texttt{
\hspace*{3.20cm} higher\_order\_derivative k f t)
}}}
\end{definition}

The function \texttt{differ\_equat\_order\_n} accepts the order of the linear differential equation \texttt{n}, list of coefficients \texttt{lst}, a differentiable function \texttt{f} and the differentiation variable \texttt{t} and returns the linear differential equation of order $n$.

Now, we model the dynamical behaviour of the activated expression as the following \holl~function:

\begin{definition}
\label{DEF:dyn_model_activ_express}
\emph{Dynamical Model of Activated Expression} \\{
\textup{\texttt{
$\vdash_{\mathit{def}}$ $\forall$$\mathtt{\alpha}$. \kww{olst\_de\_ae} $\mathtt{\alpha}$ = [Cx $\mathtt{\alpha}$; Cx (\&1)]    \\
}}
\textup{\texttt{
$\vdash_{\mathit{def}}$ $\forall$$\mathtt{{\gamma_A}^{\ast}}$. \kww{ilst\_de\_ae} $\mathtt{{\gamma_A}^{\ast}}$ = [Cx $\mathtt{{\gamma_A}^{\ast}}$]    \\
}}
\textup{\texttt{
$\vdash_{\mathit{def}}$  \kww{differ\_equat\_ae} u y t $\mathtt{\alpha}$ $\mathtt{{\gamma_A}^{\ast}}$ $\Leftrightarrow$
}} \\
\textup{\texttt{
\hspace*{0.55cm} differ\_equat\_order\_n 1 (olst\_de\_ae $\mathtt{\alpha}$) y t =
}}  \\
\textup{\texttt{
\hspace*{0.55cm}  differ\_equat\_order\_n 0 (ilst\_de\_ae $\mathtt{{\gamma_A}^{\ast}}$) u t
}}}
\end{definition}

To formally verify the transfer function of the activated expression based on its dynamical model, we first model its block diagram representation using our formalization in \holl~as:

\begin{definition}
\label{DEF:block_diagram_rep_activ_express}
\emph{Block Diagram Representation of Activated Expression} \\{
\textup{\texttt{
$\vdash_{\mathit{def}}$ $\forall$$\mathtt{\alpha}$ $\mathtt{{\gamma_A}^{\ast}}$. \kww{bdr\_ae} $\mathtt{\alpha}$ $\mathtt{{\gamma_A}^{\ast}}$ = series\_comp $\left[\mathtt{{\gamma_A}^{\ast}};\ \mathtt{\frac{Cx (\&1)}{s\ +\ Cx\ \alpha}}\right]$    \\
}}
}
\end{definition}

Next, we verify the transfer function of the activated expression based on its block diagram representation as the following \holl~theorem:

\begin{theorem}
\label{THM:block_diagram_imp_transfer_fun}
\emph{Transfer Function of Activated Expression} \\{
\textup{\texttt{
$\vdash_{\mathit{thm}}$ $\forall$$\mathtt{\alpha}$ $\mathtt{{\gamma_A}^{\ast}}$.\ \kww{[A]:} $\mathtt{(s\ +\ Cx\ \alpha)}$ $\neq$ Cx (\&0)  \vspace*{0.1cm}     \\
}}
\textup{\texttt{
 \hspace*{1.5cm} $\Rightarrow$ bdr\_ae $\mathtt{\alpha}$ $\mathtt{{\gamma_A}^{\ast}}$ = $\mathtt{\dfrac{{\gamma_A}^{\ast}}{s\ +\ Cx\ \alpha}}$     \\
}}
}
\end{theorem}

The proof of the above theorem is based on Definition~\ref{DEF:series_rep_comp} along with some arithmetic reasoning. Now, we formally verify the transfer function, obtained from Theorem~\ref{THM:block_diagram_imp_transfer_fun} based on the dynamical model as follows:

\begin{theorem}
\label{THM:dyn_model_imp_transfer_fun}
\emph{Dynamical Model Implies Transfer Function} \\{
\textup{\texttt{
$\vdash_{\mathit{thm}}$ $\forall$$\mathtt{\alpha}$ $\mathtt{{\gamma_A}^{\ast}}$ y u s.
}}  \\
\textup{\texttt{
 \hspace*{0.4cm} $\mathbf{\mathtt{\kww{[A_1]:}}}$ \&0 < $\mathtt{{\gamma_A}^{\ast}}$ $\wedge$
 \hspace*{1.0cm}  $\mathbf{\mathtt{\kww{[A_2]:}}}$ \&0 < $\mathtt{\alpha}$ $\wedge$
}}  \\
\textup{\texttt{
 \hspace*{0.4cm} $\mathbf{\mathtt{\kww{[A_3]:}}}$ $\forall$t. differentiable\_higher\_deriv u y t $\wedge$
}}  \\
\textup{\texttt{
 \hspace*{0.4cm} $\mathbf{\mathtt{\kww{[A_4]:}}}$  laplace\_exists\_higher\_deriv u y $\wedge$
}}  \\
\textup{\texttt{
 \hspace*{0.4cm} $\mathbf{\mathtt{\kww{[A_5]:}}}$ zero\_init\_cond u y $\wedge$
}}  \\
\textup{\texttt{
 \hspace*{0.4cm} $\mathbf{\mathtt{\kww{[A_6]:}}}$ ($\forall$t. differ\_equat\_ae u y t $\mathtt{\alpha}$ $\mathtt{{\gamma_A}^{\ast}}$) $\wedge$
}}  \\
\textup{\texttt{
 \hspace*{0.4cm} $\mathbf{\mathtt{\kww{[A_7]:}}}$  (laplace\_transform u s $\neq$ Cx (\&0)) $\wedge$
}} \vspace*{0.1cm} \\
\textup{\texttt{
 \hspace*{0.4cm} $\mathbf{\mathtt{\kww{[A_8]:}}}$  $\mathtt{\left(\dfrac{Cx (\&1)}{s\ +\ Cx\ \alpha}\ \neq\ Cx (\&0)\right)}$
}}  \vspace*{0.1cm} \\
\textup{\texttt{
 \hspace*{0.8cm} $\Rightarrow$ $\mathtt{\dfrac{laplace\_transform\ y\ s}{laplace\_transform\ u\ s}}$ = $\mathtt{\dfrac{\mathtt{{\gamma_A}^{\ast}}}{s\ +\ Cx\ \alpha}}$
}}
}
\end{theorem}

Assumptions $\mathtt{A_1}$-$\mathtt{A_2}$ model the positivity conditions on circuit's parameters. Assumptions $\mathtt{A_3}$-$\mathtt{A_4}$  provide the differentiability and condition of the existence of the Laplace transform of the higher-order derivative of \texttt{y} up to order $1$ and and the function \texttt{u}, respectively. Similarly, Assumption $\mathtt{A_5}$ models the zero initial conditions for \texttt{y} and \texttt{u}. Assumption $\mathtt{A_6}$ presents the dynamical behaviour of the activated expression. Assumptions $\mathtt{A_7}$-$\mathtt{A_8}$ ensure that the denominator of the transfer function, presented in the conclusion of the above theorem, provides a valid expression. Finally, the conclusion provides the transfer function of the activated expression. The proof of the above theorem is done almost automatically using the automatic tactic \texttt{TF\_TAC}, which is developed in our proposed formalization.

\subsection{Repressed Expression of Protein}\label{SUBSEC:repres_express_protein}

Figure~\ref{FIG:genetic_circuit_bd}(b) depicts the genetic circuit of the repressed expression of protein. It involves the interaction of the incoming repressing TF A with its promoter and the
regulation of the expression of gene $Y$ producing the protein Y. The block diagram representation of the repressed expression is shown in Figure~\ref{FIG:genetic_circuit_bd}(c), by a gain block of $-\gamma_x^{\ast}$ with $x = R$.
The dynamical model of the repressed expression of protein is mathematically expressed as:

\begin{equation}\label{EQ:dyn_mod_repres_expr}
\frac{dy}{dt} + \alpha y = - \gamma_R^{\ast} u
\end{equation}

The corresponding transfer function is represented as:

\begin{equation}\label{EQ:trans_fun_repres_expr}
\frac{Y (s)}{U (s)} = \frac{- \gamma_R^{\ast}}{s + \alpha}
\end{equation}

We formally verified the block diagram representation of repressed expression, its transfer function based on its block diagram and its dynamical model and the details about this verification can be found at~\cite{adnan19fabcholtp}.

Due to the undecidable nature of the higher-order logic, the verification results presented above involved manual interventions and human guidance. However, we developed the tactic \texttt{TF\_TAC}~\cite{adnan19fabcholtp} to automate the verification of the transfer functions of the biological circuits.
The distinguishing feature of our formal analysis is that all of the verified theorems are of generic nature, i.e., all of the functions and variables are universally quantified and thus can be specialized based on the requirement of analyzing the biological circuits. Whereas, in the case of computer based simulations and numerical methods, we need to model each case individually. Moreover, the inherent soundness of the theorem proving method ensures that all the required assumptions are explicitly present along with the theorem. Similarly, due to the high expressiveness of the higher-order logic, our approach allows us to model the dynamics of the biological circuits involving differential and derivative (Equations~(\ref{EQ:dyn_mod_activ_expr}),~(\ref{EQ:dyn_mod_repres_expr})) in their true form, whereas, in their model checking based analysis~\cite{yordanov2012experimentally} they are discretized and modeled using a state-transition system, which may compromise the accuracy and completeness of the corresponding analysis.


\section{Summary}\label{SEC:summary}

In this paper, we proposed a higher-order-logic theorem proving based framework for
analyzing the dynamical behaviour of the biological circuits. We formalized the dynamical model and the block diagram representations of the biological circuits, such as the activated and repressed expressions of the protein. Finally, we formally verified their transfer functions based on their dynamical models and their associated block diagram representations.
In future, we aim to formally verify some more control systems properties of the biological circuits, such as, sensitivity and stability etc.

\begin{acks}
This work was supported and funded by Kuwait University, Research
Project No. (EO 01/18).
\end{acks}

\bibliographystyle{ACM-Reference-Format}
\bibliography{sample-bibliography}

\end{document}